%% file: paper.tex
\documentclass{sig-alternate-10pt}
\usepackage{graphicx}
\usepackage{amssymb}
\usepackage{amsmath}
\usepackage{multicol,multirow}
\usepackage{epsf,psfrag,pstricks}
\usepackage{rotating,subfigure}
\newcommand{\eg}{{\it e.g.}}
\newcommand{\ie}{{\it i.e.}}
\newcommand{\etc}{{\it etc.}}

\begin{document}
\title{Deep Diving into BitTorrent Locality \\ \vspace{15pt} {\large [Please cite the IEEE INFOCOM'11 version of this paper]}}

\numberofauthors{3}
\author{
\alignauthor
Ruben Cuevas\\
\affaddr{Univ. Carlos III de Madrid}\\
{\tt rcuevas@it.uc3m.es}
\alignauthor
Nikolaos Laoutaris\\
\affaddr{Telefonica Research}\\
{\tt nikos@tid.es}
\alignauthor
Xiaoyuan Yang\\
\affaddr{Telefonica Research}\\
{\tt yxiao@tid.es} \and
\alignauthor
Georgos Siganos\\
\affaddr{Telefonica Research}\\
{\tt georgos@tid.es}
\alignauthor
Pablo Rodriguez\\
\affaddr{Telefonica Research}\\
{\tt pablorr@tid.es}
}

\maketitle

\input{abstract}
\input{introduction}
\input{implicit}

\input{demographics}

\input{trafficmatrix}
\input{overlayconstruction}

\input{understanding}
\input{prototype}
\input{related}

\input{conclusions}

{\small
\bibliographystyle{plain}
\bibliography{Nikos_EN}}

\appendix

\input{bmatching_basic}

\input{validation}

\input{completion}

\end{document}

%% file: abstract.tex
\begin{abstract}

A substantial amount of work has recently gone into localizing
BitTorrent traffic within an ISP in order to avoid excessive and
often times unnecessary transit costs. Several architectures and
systems have been proposed and the initial results from specific
ISPs and a few torrents have been encouraging. In this work we
attempt to deepen and scale our understanding of locality and its
potential. Looking at specific ISPs, we consider tens of thousands
of concurrent torrents, and thus capture ISP-wide implications
that cannot be appreciated by looking at only a handful of
torrents. Secondly, we go beyond individual case studies and
present results for the top 100 ISPs in terms of number of users
represented in our dataset of up to 40K torrents involving more
than 3.9M concurrent peers and more than 20M in the course of a
day spread in 11K ASes. We develop scalable methodologies that
permit us to process this huge dataset and answer questions such
as: ``\emph{what is the minimum and the maximum transit traffic
reduction across hundreds of ISPs?}'', ``\emph{what are the
win-win boundaries for ISPs and their users?}'', ``\emph{what is
the maximum amount of transit traffic that can be localized
without requiring fine-grained control of inter-AS overlay
connections?}'', ``\emph{what is the impact to transit traffic
from upgrades of residential broadband speeds?}''.

\end{abstract}

%% file: introduction.tex
\section{introduction}

Most design choices in P2P applications are dictated by end user performance and implementation simplicity. Bootstrapping is one such example: a new node joins a P2P overlay by connecting to a \emph{Random} set of neighbors. This simple process provides fault tolerance and load balancing to end users and implementation simplicity to developers. Its downside, however, is that it is completely oblivious to the requirements and operating constraints of ISPs and thus it often ends up causing serious problems such as increasing the transit costs, worsening the congestion of unpaid peering links~\cite{leighton-2009-improving}, and expediting the upgrade of DSLAMs. Therefore, several ISPs have allegedly started rate limiting or blocking P2P traffic~\cite{Marcel08:bittorrentblocking}. In response, P2P applications have tried to conceal and evade discriminatory treatment by using dynamic ports and protocol encryption.

Much of this tension can be avoided by biasing the overlay construction of P2P towards \emph{Locality}. It is known that geographic proximity often correlates with overlap of consumption patterns~\cite{Karagiannis05:P2PIMC} and thus bootstrapping P2P users with other nearby ones can confine P2P traffic within ISPs instead of letting it spill to other domains over expensive transit links. This simple idea has received much attention lately since it is generic and thus can be applied to a variety of P2P applications independently of their internal logic (scheduling, routing, \etc). Systems like P4P~\cite{Xie08:P4P} and ONO~\cite{choffnes08:ONO} have been proposed for localizing the traffic of the BitTorrent file sharing protocol~\cite{Cohen03:BitTorrent}. P4P proposes bilateral cooperation between ISPs and P2P applications, whereas ONO is a client-side solution that does not involve the ISP.

Despite the interesting architectures and systems that have already been proposed, we believe that we still stand on preliminary ground in terms of our understanding of this technology. Although the main ideas are straightforward, their implications are quite the opposite, for several reasons. First, different torrents can have quite diverse \emph{demographics}: a blockbuster movie has peers around the world and thus can create much more transit traffic than a local TV show whose peers are mostly within the same country/ISP, especially if language gets in the way. Predicting the ISP-wide transit traffic due to P2P amounts to understanding the demographics of thousands of different torrents downloaded in parallel by all the customers. Things become even more complicated in the case of the BitTorrent protocol whose free-riding avoidance scheme makes peers exchange traffic predominately with other peers of similar speed~\cite{Legout07:Clustering}. Thus even if two ISPs have similar demographic composition, the fact that they offer different \emph{access speeds} can have a quite pronounced impact on the amount of transit traffic that they see. The combined effect of demographics and access speeds makes it risky to generalize observations derived from a particular ISP and few individual torrents.

\section{Our contributions}

Our works provides detailed case studies under representative ISP-wide workloads as well as holistic views across multiple (hundreds) of ISPs. In all cases we demand that the input be as representative as possible (demographics and speed of different ISPs) and the methodology be scalable without sacrificing essential BitTorrent mechanisms like the unchoke algorithm, the least replicated first chunk selection policy, and the effect of seeders. We collected representative input data by scraping up to 100K of torrents of which at least 40K had active clients from Mininova and Piratebay, the two most most popular torrent hosting sites in the world according to the Alexa Ranking. We then queried the involved trackers to construct a map of BitTorrent demand demographics of up to 3.9M concurrent users and more than 21M total users over the course of a day, spread over 11K ISPs. For all those ISPs we obtained speeds from a commercial speed-test service~\cite{Stanford:page} and from the iPlane project~\cite{iplane_osdi}.

Our datasets are too big to conduct emulation or simulation studies. To process them, we employ two scalable methodologies: a probabilistic one for deriving speed-agnostic upper and lower bounds on the number of piece exchanges that can be localized within an ISP given its demand demographics and a more accurate deterministic one that estimates the resulting traffic matrix taking into consideration the speeds of different ISPs. The probabilistic technique allows us to scale our evaluation up to as many ISPs as we like (we report on the 100 largest ones) whereas the deterministic one allows us to zoom in into particular ISPs and refine our estimation of transit traffic and end-user QoS. With these two tools, we study the performance of a rather broad family of overlay construction mechanisms that includes: \emph{Locality Only If Faster, (LOIF)}, an end-user QoS preserving overlay that switches remote neighbors for locals only when the latter are faster, and \emph{Locality}, a simple policy that maximizes transit savings by switching as many remote neighbors as possible with local ones, independently of relative speed.

\vspace{3pt}

\noindent \textbf{Summary of results:} We shed light to several yet unanswered questions about BitTorrent traffic. Specifically:

\vspace{2pt}

\begin{itemize}

\item We use the demand demographics of the 100 largest ISPs from our
dataset to derive speed agnostic upper and lower bounds on the
number of chunk exchanges that can be kept local. In half of the
ISPs, Locality keeps at least 42\% and up to
72\% of chunks internal, whereas Random can go from less than 1\% up to 10\%.

\end{itemize}

Next we focus on the three largest US and the three largest European ISPs in our dataset and derive their traffic matrices using both demographic and speed information. These detailed case studies reveal the following:

\begin{itemize}

\item LOIF preserves the QoS of users and reduces the transit traffic
of fast ISPs by around 30\% compared to Random. In slower ISPs the
savings are around 10\%.

\item Locality achieves transit traffic reductions that
peak at around 55\% in most of the ISPs that we considered.
The resulting penalty on user download rates is typically less
than 6\%.

\item The barrier on transit traffic reduction is set by ``unlocalizable'' torrents, \ie, torrents with one or very few nodes inside an ISP. Such torrents account for around 90\% of transit traffic under Locality and are requested by few users of an ISP ($\sim$10\%). In a sense, the majority of users is subsidizing the transit costs incurred by the few users with a taste for unlocalizable torrents.

\item By limiting the number of allowed inter-AS overlay links
per client huge reductions of transit ($>$95\%) are possible. The
resulting median penalty is around 20\%  but users on ``unlocalizable'' torrents incur huge reduction of QoS (99\%).

\item Finally, we show that, contrary to popular belief, increasing the speed of access connections does not necessarily keep more traffic local as it might bring an ISP within unchoke distance from other fast ISPs which previously did not send it traffic.

\end{itemize}

Overall our results show that there is great potential from locality for both ISPs and users but there also exist some cases in which locality needs to be approached with caution. Cashing in this potential in practice is a non-trivial matter, but seems to be worthy of further investigation.

The remainder of the article is structured as follows. In Sect.~\ref{sec:implicit} we derive upper and lower bounds on the number of localized unchokes under Random and Locality overlays, independently of ISP speed distributions. In Sect.~\ref{sec:demographics} we present our measurement study of BitTorrent demographics. We also define a metric for explaining the performance of Random when factoring in real speed distributions across ISPs. In Sect.~\ref{sec:bittorrenttrafficmatrix} we present a methodology for estimating BitTorrent traffic matrices and in Sect.~\ref{sec:overlayconstruction} we define the family of overlay construction policies that we use later in our study. Sect.~\ref{sec:understanding} characterizes the win-win situations and the tradeoffs between ISPs and users under different locality policies. In Sect.~\ref{sec:prototype} we present a validation prototype for studying locality using live torrents and factoring in network bottlenecks. In Sect.~\ref{sec:related} we look at related work and we conclude in Sect.~\ref{sec:conclusions}.

%% file: implicit.tex
\section{Why not a Random Overlay?}\label{sec:implicit}

Our goal in this section is to understand the cases in which a Random selection of neighbors localizes traffic well, and the ones in which it fails thereby creating the need for locality-biased neighbor selection. To do so we first need to understand the \emph{stratification effect}~\cite{Legout07:Clustering} arising due to the unchoke algorithm~\cite{Cohen03:BitTorrent} used by BitTorrent to combat free-riding. According to this algorithm, a node monitors the download rates from other peers and ``unchokes'' the $k$ peers (typically 4--5) that have provided the highest rates over the previous 20~$sec$ interval. These peers are allowed to fetch missing chunks from the local node over the next 10~$sec$ interval. Therefore, as long as there are chunks to be exchanged between neighbors (LRF chunk selection works towards that~\cite{Cohen03:BitTorrent}), peers tend to stratify and communicate predominantly with other peers of similar speed.

In this section, we employ probabilistic techniques to help us build some basic intuition on the consequences of stratification on inter-domain traffic. We focus on a single ISP $A$ and torrent $T$ and analyze the conditions under which Random localizes sufficiently within $A$ the traffic due to $T$. In Sect.~\ref{sec:demographics} we will examine the effects from multiple torrents with different demographics as well as the effect of speed differences between ISPs. In Sect.~\ref{sec:bittorrenttrafficmatrix} we will go into an even more accurate model that captures more precisely the unchoke behavior of leechers (including optimistic unchokes) as well as the different behavior of seeders. We will also discuss the impacts of torrents that are not in steady-state and develop a model for them (Appendix~\ref{appendix:bmatching_completion}).

\subsection{Sparse mode -- the easy case for Random}

Let $V(T)$ denote the set of BitTorrent nodes participating in $T$, and $V(A,T)\subseteq V(T)$ the subset that belongs to ISP $A$. We say that ISP $A$ is on \emph{sparse mode} with respect to torrent $T$ if the nodes outside $A$ that participate in $T$ have very dissimilar speeds with nodes that are within $A$. In this case, because of stratification, local nodes of $A$ will talk exclusively to each other irrespectively of other remote nodes in their neighborhood. Then to confine all unchokes within $A$, each local node needs to know at least $k$ other local neighbors. If $W$ denotes the size of a neighborhood (40 upon bootstrap and growing later with incoming connections), then for Random to localize all traffic it has to be that a random draw of $W$ out of the total $|V(T)|-1$ (-1 to exclude the node that is selecting) nodes yields at least $k$ local ones. The probability of getting $x$ ``successes'' (\ie, local nodes) when drawing randomly $W$ samples from a pool of $|V(T)|-1$ items, out of which $|V(A,T)|-1$ are ``successes'' is given by the Hyper-Geometric distribution $\mathrm{HyperGeo}(x,|V(T)|-1,|V(A,T)|-1,W)$~\cite{Feller1968:probability}. Thus the expected number of localized unchokes is

\begin{equation}\label{eq:localunchokesrandomsparse}
\small
\sum_{x=0}^{\min(|V(A,T)|-1,W)} \hspace{-20pt} \min(x,k)\cdot \mathrm{HyperGeo}(x,|V(T)|-1,|V(A,T)|-1,W)
\end{equation}

Taking the mean value of the distribution we can write a condition for Random to localize well in sparse mode:

\begin{equation}\label{eq:condsparse}
\small
\frac{W\cdot(|V(A,T)|-1)}{|V(T)|-1}\geq k
\end{equation}

\subsection{Dense mode -- things getting harder}

ISP $A$ is on \emph{dense mode} with respect to $T$ if the remote nodes participating in $T$ have similar speeds to the nodes of $A$. In this case stratification does not automatically localize traffic inside $A$. From the standpoint of the unchoke algorithm, both local and remote nodes look equally good and thus the number of localized unchokes depends on their ratio in the neighborhood. Thus, although in sparse mode a random draw yielding $x\leq k$ local nodes would keep all $x$ unchokes local, in dense mode it keeps only $k \cdot x/W$ of them local in expectation. To get the expected number of localized unchokes in dense mode we have to substitute $\min(x,k)$ with $k\cdot x/W$ in Eq.~(\ref{eq:localunchokesrandomsparse}).

\subsection{The promise of Locality}

Let's now consider Locality, an omniscient overlay construction mechanism that knows all local nodes and thus constructs highly localized neighborhoods by providing each node with as many local neighbors as possible, padding with additional remote ones only if the locals are less than $W$. Then in sparse mode Locality localize all unchokes as long as $|V(A,T)|-1\geq k$, which is a much easier condition to satisfy than the one of Eq.~(\ref{eq:condsparse}), else it localizes only $|V(A,T)|-1$. In dense mode Locality localizes all unchokes as long as $|V(A,T)|-1\geq W$.

\subsection{Locality gains are higher in dense mode}

Overall Random localizes sufficiently in sparse mode as long as it can get a small number of local nodes in each neighborhood. In dense mode things become more challenging as it no longer suffices to guarantee a small threshold of locals but instead Random has to have a strong majority of locals in each neighborhood. In both modes, Locality has to satisfy easier conditions to localize the same number of unchokes. Further, we can actually prove that the improvement factor of Locality over Random in terms of the number of localized unchokes is higher in dense mode than in sparse mode. We consider only the case with $|V(A,T)|-1\geq k$ and $|V(T)|-1\geq W$ (the other ones can be worked out similarly). Based on the previous analysis we get that the expected improvement factor in sparse mode is:

\begin{equation}\label{eq:factorsparse}
\small
\frac{k}{W\cdot\frac{|V(A,T)|-1}{|V(T)|-1}}
\end{equation}

In dense mode for $|V(A,T)|-1\geq W$ the improvement factor is:

\[
\small
\frac{k}{k\cdot\frac{|V(A,T)|-1}{|V(T)|-1}}
\]

which is greater than Eq.~(\ref{eq:factorsparse}) since $W>k$. For $|V(A,T)|-1< W$ the improvement factor is:

\[
\small
\frac{k \cdot \frac{|V(A,T)|-1}{W}}{k\cdot\frac{|V(A,T)|-1}{|V(T)|-1}}=\frac{|V(T)|-1}{W}
\]

which can be checked to be greater than Eq.~(\ref{eq:factorsparse}) for the case with $|V(A,T)|-1\geq k$. In Sect.~\ref{subsec:bounds} we will use measurement to determine the ISP-wide improvement factor.

%% file: demographics.tex
\section{Demographics of BitTorrent}\label{sec:demographics}

We conducted a large measurement study of BitTorrent demand
demographics. We begin with a presentation of our
measurement methodology and then use the obtained demographics to
derive upper and lower bounds on the number of localized regular unchokes
under Random and Locality. At the end of the section we
incorporate the effect of speed differences among ISPs and show
that it is non trivial to predict what happens to the transit
traffic of an ISP when it upgrades the speed of its residential
accesses.

\subsection{Measurement methodology}\label{subsec:measurementmethodology}

We developed a custom BitTorrent crawler that obtains a
snapshot of the IP addresses of all the clients participating in a
set of torrents that are provided as input. In
Table~\ref{table:inputtorrentsets} we present the different sets
of torrents used in our study. Our crawler first scrapes a torrent
indexing site to obtain {\tt .torrent} meta information files.
From them it obtains the addresses of the corresponding trackers.
It queries repeatedly the trackers and uses the gossip protocol implemented in the latest versions of BitTorrent to obtain all the IP addresses
of clients participating in each torrent. The gathered IP addresses are
mapped to ISPs and countries using the MaxMind database
\cite{maxmind:page}. The crawler also obtains the
number of seeders and leechers in each torrent. Crawling an individual torrent 
takes less than 2 minutes. Thus we get a
pretty accurate ``snapshot'' of each torrent, \ie, we are sure
that the obtained IPs are indeed present at the same time.
The time difference between the first and last crawled
torrent was up to 90 minutes for the largest dataset ({\tt mn40K}).
However, we tracked individual torrent populations and found them
to be quite stable across a few hours. Thus our dataset is
similar to what we would get if we used a very large number of
machines to crawl more torrents in parallel. We specifically
wanted to avoid that since it would be a denial of service attack against the hosting sites and the trackers.

\begin{table}[tb]
\begin{center}
\begin{tabular}{ccccc}
\hline
Set name & Source & Torrents & \# IPs & \# ISPs\\  \hline
{\tt mn40K} & Mininova & latest 40K & 3.9M & 10.4K\\ %\hline
{\tt mn3K} & Mininova & latest 3K & 17.4M & 10.5K\\ %\hline
{\tt pb600} & Piratebay & 600 most popular & 21.9M & 11.1K\\ \hline
\end{tabular}
\end{center}
\caption{Torrent sets collected in the period Aug-Oct 2009. For
{\tt mn40K} we collected three versions, with one week in between
them. We actually crawled 100K torrents but only around 40K had
peers. For {\tt mn3K} and {\tt pb600} we repeated the crawl
every hour for one day. The \#IPs and \#ISPs for {\tt mn40K} are
per snapshot, whereas for {\tt mn3K} and {\tt pb600} are daily
totals.} \label{table:inputtorrentsets}
\end{table}

\begin{figure}[tb]
\centering
\includegraphics[width=3.2in]{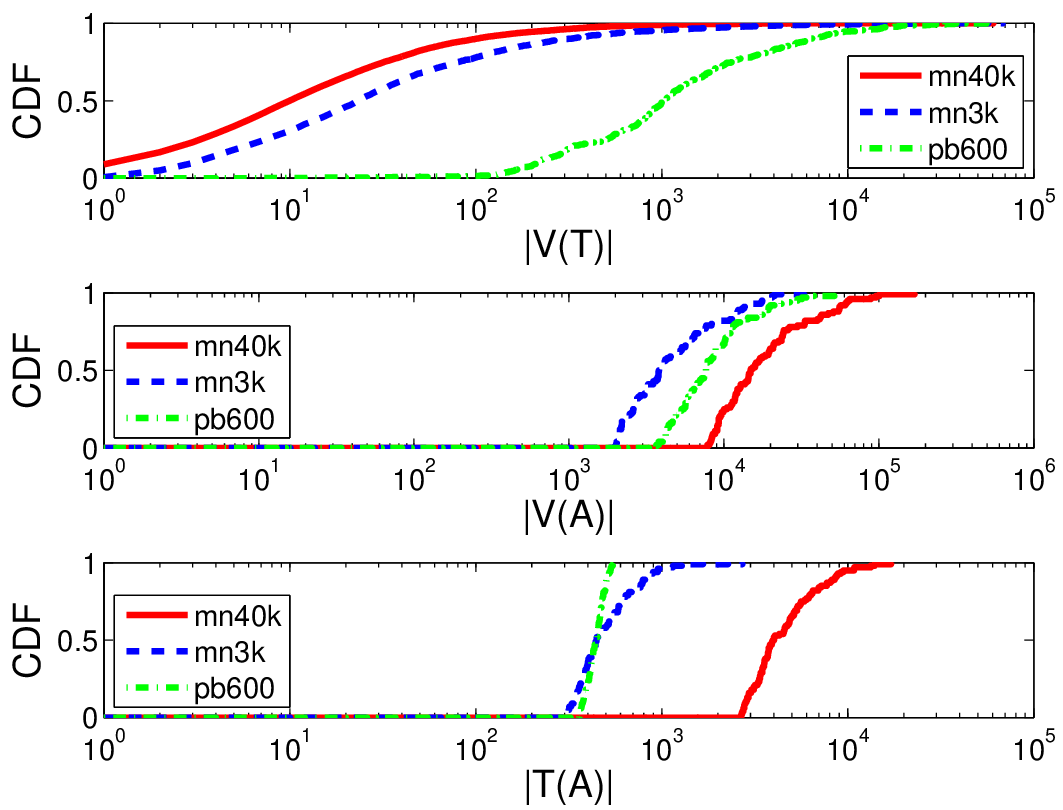}\hfill
\caption{Summary statistics for the measured BitTorrent
demographics. Cdfs for: $|V(T)|$, the number of clients in a
torrent, $|V(A)|$, the total number of clients in
an ISP across all its torrents, and $|T(A)|$, the number of distinct torrents requested by
the clients of an ISP.} \label{fig:measurementsummary}
\end{figure}

\subsection{High level characterization of the dataset}

We use the following definitions. We let $\mathcal{T}$ denote the
set of torrents appearing in our measurements and $\mathcal{A}$
the set of ISPs that have clients in any of the
torrents of $\mathcal{T}$. We let $T(A)$ denote the set of
torrents that have at least one active client in $A$, and
$V(A)=\bigcup_{T\in T(A)} V(A,T)$ the set of all clients of $A$
participating in any of the torrents $T(A)$. In
Fig.~\ref{fig:measurementsummary} we summarize the measured
demographics. Some points worth noting:

The largest torrent has approximately 60K clients in all three
datasets. Looking at the large set, {\tt mn40K} we see that most
torrents are small, as has already been shown~\cite{Piatek09:locality,Menasche09:bundling}. {\tt
mn3K} has relatively bigger torrents since it is a subset of
most recent torrents of {\tt mn40K}, and recency correlates with
size. {\tt pb600} holds by definition only big torrents.

Looking at the number of peers and torrents per ISP we see that
{\tt mn40K} has bigger values which is expected since it is a much
bigger dataset than the other two and thus contains more and
bigger ISPs (notice that in
Table~\ref{table:inputtorrentsets} the numbers for {\tt mn40K} are
per snapshot, whereas for the other two are aggregates over a day,
\ie, totals from 24 snapshots).

\subsection{Speed agnostic bounds for the measured demand demographics}\label{subsec:bounds}

\begin{figure}[tb]
\centering
\includegraphics[width=3.3in]{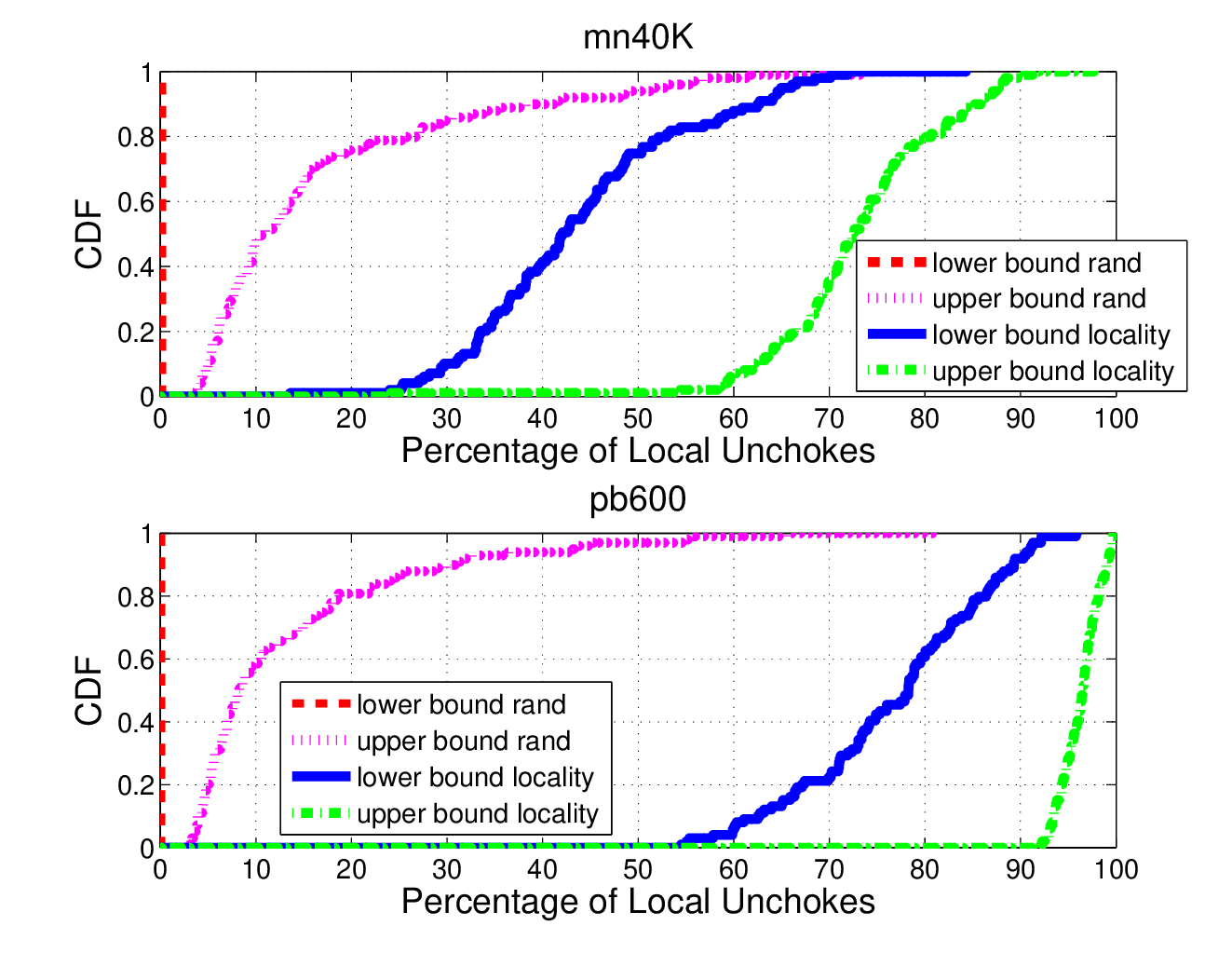}\hfill
\caption{CDF of the upper and lower bound on the number of localized unchokes under Random and Locality for top-100 ISPs in number of clients. Top: {\tt mn40K} dataset. Bottom: {\tt pb600} dataset.}
\label{fig:sparse_dense_metrics}
\end{figure}

In Sect.~\ref{sec:implicit} we defined the notions of sparseness
and denseness for one ISP and a single torrent and noted that
sparseness helps to localize traffic whereas denseness makes it
harder. Therefore, by assuming that all the torrents $T(A)$
downloaded in $A$ are concurrently in sparse mode we can get an
upper bound on the expected number of unchokes that can be localized by an
overlay construction policy for the given demand demographics and
any speed distribution among different ISPs. Similarly, by
assuming that all torrents are in dense mode we get a lower bound.
In Fig.~\ref{fig:sparse_dense_metrics} (top) we plot the upper
and lower bound on localized unchokes for Random and Locality for the top-100 ISPs in number of clients in the {\tt mn40K} dataset (this list includes all major ISPs and amounts for more than 68\% of all IPs in the dataset). These bounds were computed using formula~(\ref{eq:localunchokesrandomsparse}) and its corresponding version for dense mode for single torrents and iterating over all $T\in T(A)$ from
our demographics dataset adding each contribution with weight
$|V(A,T)|/\sum_{T'\in T(A)} |V(A,T')|$ to capture the relative
importance of $T$ for $A$.

%Since the majority of ISPs have few
%clients in our dataset, we present cdfs only for the top-100
%largest ISPs in terms of number of IPs.

The lower bound for Random is very close to 0. This happens because for the huge majority of torrents, an ISP has only a small minority of the total nodes in the torrent. In dense mode, Random needs to get most of these few locals with a random draw which is an event of very small probability. On the other hand, this small minority of nodes performs much better in sparse mode yielding an upper bound for Random that is at least 10.94\% in half of the top-100 ISPs. Locality has strikingly better performance. Its lower bound is at least 42.35\% and its upper bound 72.51\% in half of the top-100 ISPs. The huge improvement comes from the fact that Locality requires the mere existence of few local nodes in order to keep most unchokes inside an ISP. As noted earlier, the improvement factor is greater in the difficult case (the lower bound goes from 0 to above 42\% in half of the cases) while it is also quite big in the easy case (improvement factor of at least 6.63 in half of the cases).

In Fig.~\ref{fig:sparse_dense_metrics} (bottom) we recompute these bounds based on the {\tt pb600} dataset. In this case, the upper bound of Random is lower since nodes from the same ISP become an even smaller minority in very large torrents. On the other hand, Locality benefits in terms of both upper and lower bounds. This happens because the bounds for Locality, unlike Random, depend on the absolute rather than the relative number of local nodes, which increases with larger torrents. These bounds paint, to the best of our knowledge, the most extensive picture reported up to now in terms of covered ISPs and torrents of the potential of locality given the constraints set by real demand
demographics.

\begin{figure}[tb]
\centering
\includegraphics[width=3.3in]{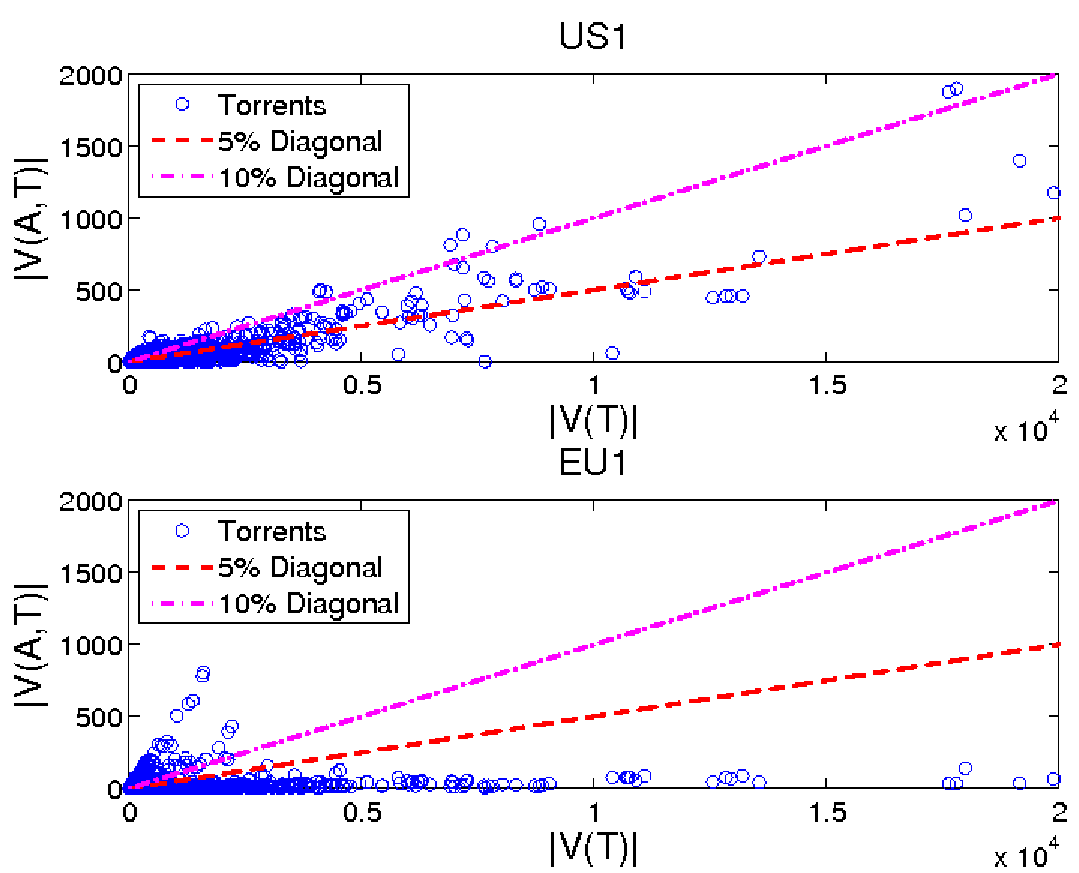}
\caption{Nodes in ISP A, $|V(A,T)|$, vs. total torrent size, $|V(T)|$, for US1 (top) and EU1 (bottom).}
\label{fig:demographics_casestudy}
\end{figure}

\subsection{Factoring the effect of speed}\label{subsec:inherentdefinition}

The notions of sparseness and denseness have been useful in deriving speed-agnostic performance bounds based on the demand demographics and the overlay construction policy. To refine our analysis and answer more detailed questions we turn our attention now to the effect of speed. We do so through what we call \emph{Inherent Localizability}.

% that will help us understand what happens in reality to a torrent when the number of remote nodes with similar speed with the local ones depends on the demographics and the speed distribution between ISPs, while it is also varying from torrent to torrent. With this metric we will get a more precise feel than with the previous bounds about the number of unchokes that can be localized in each case.

%\subsubsection{Inherent localizability of an ISP}

Let $A(T)$ denote the set of ISPs that have clients in torrent $T$. Let also $U(A)$ denote the uplink speed of nodes in ISP $A$. We focus on the uplink speeds because they are typically the bottleneck in highly asymmetric residential broadband accesses~\cite{Dischinger07:ResidentialBroadband}. For now it suffices to assume that speeds differ only between ISPs (we relax this in Sect.~\ref{sec:bittorrenttrafficmatrix} by considering speed distributions within an ISP). We define the \emph{Inherent Localizability} $I_q(A,T)$ of torrent $T$ in ISP $A$ as follows:

\[
\small
I_q(A,T)=\frac{|V(A,T)|}{\sum_{A'\in A(T)} |V(A',T)|\cdot \mathcal{I}(A,A',q) },
\]

where, $\mathcal{I}(A,A',q)=1$ iff $U(A)\cdot (1-q)\leq U(A')\leq U(A)\cdot (1+q)$, and 0 otherwise. The parameter $q\in[0,1]$, captures the maximum speed difference that still allows a local node of $A$ and a remote node of $A'$ to unchoke each other. In reality $q$ can be arbitrarily large since very fast nodes can unchoke much slower ones in the absence of other fast nodes. We use this simple metric here in order to gain a basic intuition into the combined effects of speed and demographics and discard it later in the context of a more accurate albeit more complex model for predicting unchoke decisions in Sect.~\ref{sec:bittorrenttrafficmatrix}. The inherent localizability $I_q(A)$ of ISP $A$ across all its torrents is simply the weighted sum by $|V(A,T)|/|V(A)|$ of its $I_q(A,T)$'s for all torrents it participates in. $I_q(A)$ captures the density of $A$'s nodes in torrents that it shares with other ISPs that have similar speed. Due to stratification, unchokes will take place among those nodes. For Random, $I_q(A)$ determines completely its ability to localize unchokes. $I_q(A)$ also impacts on Locality. However, Locality's overall performance depends on the absolute number of local peers.

\subsection{Does being faster help in localizing better?}

In this section we use inherent localizibility to study the effect of access speed on the ability of Random to keep unchokes internally in an ISP. ISPs have a natural interest in this question since on one hand they want to upgrade their residential broadband connections to fiber but on the other hand, they wonder how this will impact their transit and peering traffic. Next we present a case study showing that it is difficult to come up with such predictions without using detailed demographic/speed information and corresponding methodologies to capture their combined effect.

\subsubsection{A European and an American ISP}\label{subsubsection:casestudy1}

\begin{figure}[tb]
\centering
\includegraphics[width=3.3in]{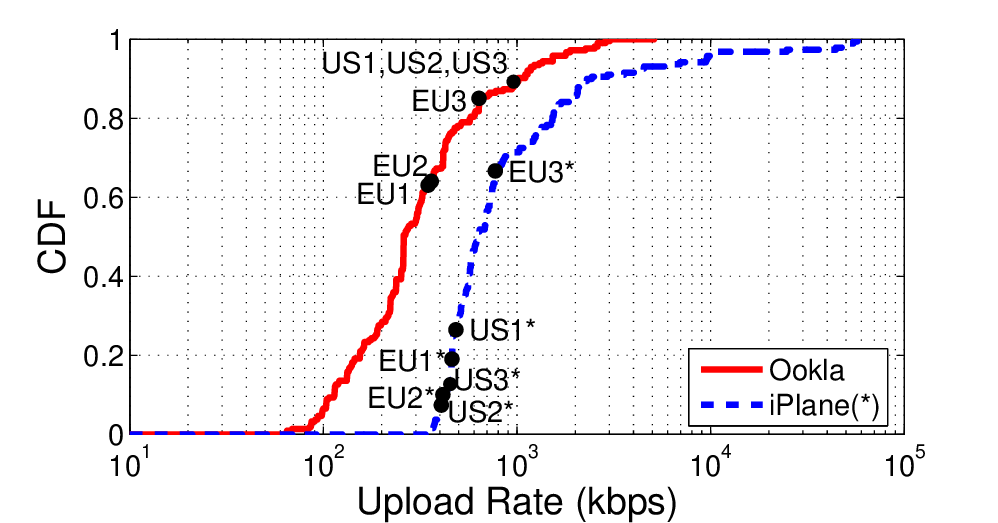}\hfill
\caption{CDF of uplink speeds per country. EU1--EU3, US1--US3 are ISPs studied in Sect.~\ref{sec:understanding}.}
\label{fig:cdf_country_speed}
\end{figure}

Consider the following two ISPs from our dataset {\tt mn40K}: US1, with the largest population of nodes in America (according to our different datasets) and median upload speed 960 Kbps, and EU1, with the largest population of nodes in Europe and median upload speed 347 Kbps. In Fig.~\ref{fig:demographics_casestudy} we plot $|V(A,T)|$ vs. $|V(T)|$ for all $T\in T(A)$  for the two ISPs. A quick glance at the figure reveals that the two ISPs are radically different in terms of demand demographics. Because of the proliferation of English and its English content, US1 is participating in globally popular torrents. In the figure, the US1 torrents that are globally large (high $|V(T)|$) have a lot of clients inside US1. Also, torrents that are popular in US1 are also globally popular. In EU1 the picture is very different. The largest torrents inside EU1 are not among the largest global ones, whereas only very few globally popular torrents are also popular inside EU1. This has to do with the fact that EU1 is in a large non-English speaking European country that produces and consumes a lot of local, or locally adapted content.

\begin{figure}[tb]
\centering
\includegraphics[width=3.3in]{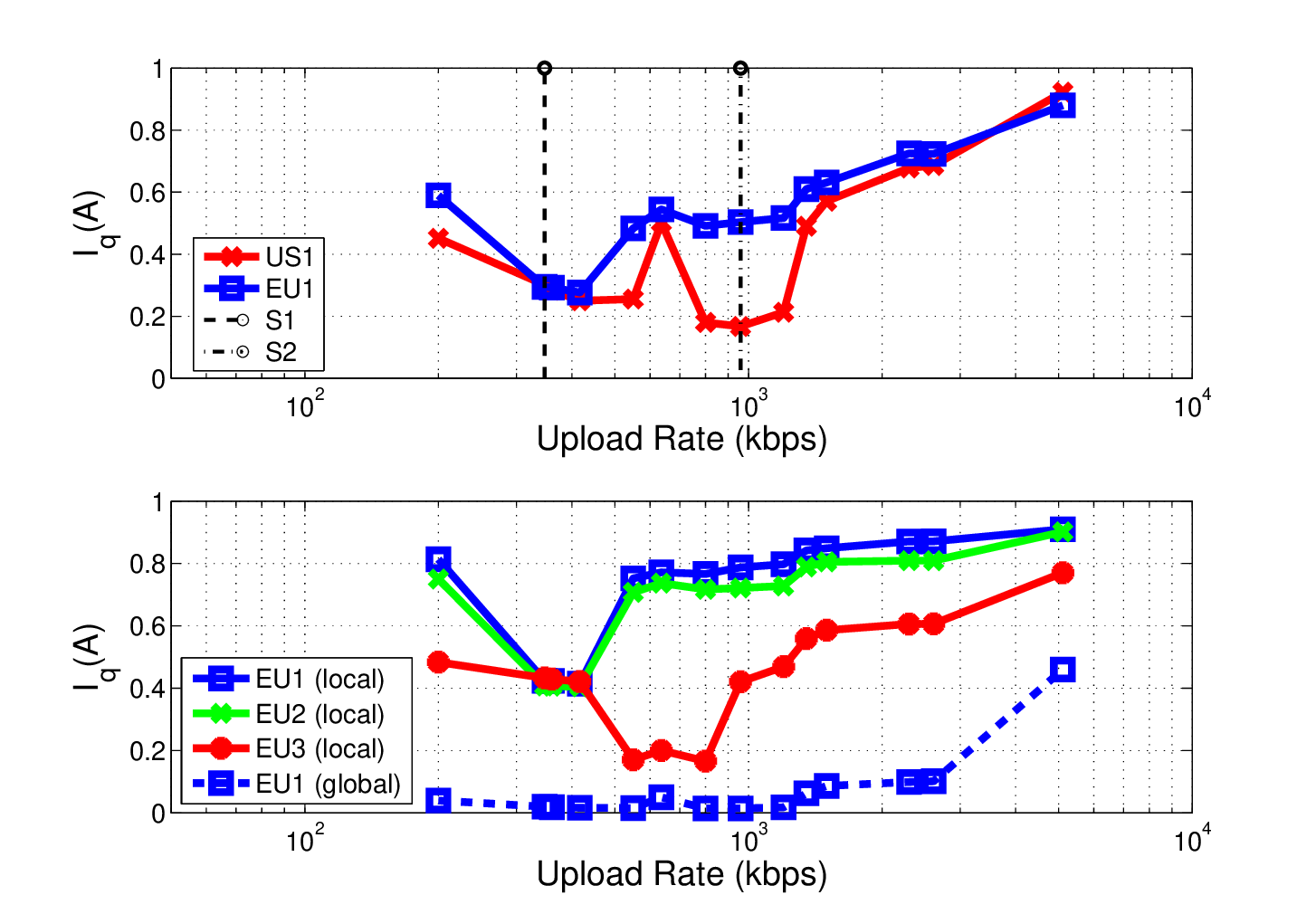}
\caption{Top: the inherent localizability of US1 and EU1 for different speeds based on all their torrents. Bottom: The inherent localizability of 3 European ISPs based on their 10 most popular local torrents and the 10 most popular torrents across the entire dataset.}
\label{fig:localaizability}
\end{figure}

\vspace{5pt}
\subsubsection{The impact of demographics and speed on inherent localizability }

We will now compute the inherent localizability of EU1 and US1. To do this we need the speeds $U(A)$ for all $A$ that participate in common torrents with the two ISPs. We have obtained these speeds from the Ookla Speedtest service~\cite{Stanford:page}. This data set includes measurements of both upload and download speeds of over 19 million IP client addresses around the world. In Fig.~\ref{fig:cdf_country_speed} we plot the cdf of median country speed based on the above dataset. It is interesting to observe that almost 80\% of the countries have similar speeds that are below 610 Kbps where the few remaining ones are sparsely spread in the range from 610 Kbps to 5.11 Mbps. We also plot the corresponding cdf from iPlane~\cite{iplane_osdi} which we use later for validation.

Using the above demographics and speeds we plot in Fig.~\ref{fig:localaizability} (top) the localizability of ISP $A\in$\{EU1,US1\} for different $U(A)$, \ie, we plot how the localizability of the two ISPs would change if we changed their speeds while keeping the speeds of all other ISPs fixed. We have assumed $q=0.25$. Results are similar for most $q<0.5$ whereas for larger ones speed starts becoming marginalized because high $q$'s imply that any node can unchoke any other one. There are two points to keep from this figure. First, the localizability of EU1 is generally higher than that of US1 for the same speed. This means that if the two ISPs had similar speed, then the demographic profile of EU1 depicted earlier in Fig.~\ref{fig:demographics_casestudy} would lead to a higher inherent localizability since this ISP holds a larger proportion of the content requested by its users. Thus Random would perform better in EU1 than in US1.

A second point to notice is that $I_{0.25}(A)$ is changing non-monotonically with $U(A)$. This happens because the set of remote ISPs and consequently the number of remote clients that can be unchoked by clients of $A$ due to similar speed (within the margins allowed by a given $q$) changes as we vary the speed of $A$. If the torrents were spread uniformly across all the ISPs, and ISPs had similar size, then due to the sparsification of ISPs on the high speed region (Fig.~\ref{fig:cdf_country_speed}), $I_{0.25}(A)$ would increase monotonically with $U(A)$. The real demographics and sizes of ISPs, though, lead to the depicted non-monotonic behavior that exhibits only a general trend towards higher intrinsic localizibility with higher local speed. This has important consequences on the expected amount of transit traffic under different speeds. For example, by going from speed S1 = 347 Kbps to S2 = 960 Kbps, the inherent localizability of EU1 increases from around 0.3 to around 0.5 and as a consequence its transit traffic under Random would decrease as more unchokes would stay inside the ISP. The opposite however happens for US1. Increasing the speed from S1 to S2 reduces the inherent localizability from 0.3 to 0.2, effectively increasing the number of unchokes going to remote nodes and thus the transit traffic as well.

\subsubsection{Local and global torrents}

It seems very difficult to devise simple rules of thumb for predicting how $I_q(A)$ will change with $U(A)$ without using detailed demographic and speed information as we did earlier. The complication owes to the torrent mix of each ISP, which includes both \emph{global} and \emph{local} torrents. Global torrents are those very popular torrents consumed by users around the word. Global torrents are omnipresent in the entire speed range, but since the country speed cdf sparsifies at higher ranges (Fig.~\ref{fig:cdf_country_speed}), fewer of them will be encountered as remote neighbors when an ISP upgrades to such speeds. This leads to more internal unchokes of global torrents, effectively making the inherent localizibility of global torrents a monotonic function of speed.

Local torrents exist at specific ISPs and speed ranges and thus their behavior during speed upgrades is more difficult to predict. For example, an ISP at a French speaking African country will see its unchokes to and from remote ISPs increasing if it upgrades its residential accesses to speeds that bring it near the offered speeds in France, Belgium, and Canada. If it upgrades to even faster speeds though, its remote unchokes will fall since its local users would rather unchoke each other than peers in other countries. In Fig.~\ref{fig:localaizability} (top) the localizability of US1 fell at around 1Mbps because it entered the region of many other US ISPs and thus started exchanging unchokes with them for content that although in English, is local to US (local TV, music, \etc). In Fig.~\ref{fig:localaizability} (bottom) we compute the inherent localizability of the 10 most popular local torrents in 3 European countries and the corresponding 10 most popular across the entire dataset. The global torrents change monotonically whereas local ones do not. The main point here is that since the interplay between speed and demographics is complicated, an ISP can use our methodology to actually obtain an informed prediction of the impact of planned changes to its residential broadband offerings on its transit traffic.

%% file: trafficmatrix.tex
\section{BitTorrent traffic matrices}\label{sec:bittorrenttrafficmatrix}

Our analysis up to now has been used for building up a basic intuition about the parameters that affect the performance of Random and Locality. However it has a number of shortcomings. First, it makes the simplifying assumption that nodes whose speeds do not differ more than a multiplicative factor $(1\pm q)$ unchoke each other. The problem with this assumption is that depending on the speed distribution, there may not be a single $q$ that predicts all the unchokes. Also, the analysis does not capture the behavior of seeders or the optimistic unchokes from leechers. In the next section we develop a more accurate model that addresses all these shortcomings and predicts the actual traffic matrix resulting from a set of torrents. Our objective is to estimate the aggregate amount of traffic routed to an ISP transit link due to the torrents of our demographic datasets of Table~\ref{table:inputtorrentsets}.

We start with fast numeric methods that capture the unchoking behavior in steady-state, \ie, when the Least Replicated First (LRF) chunk selection algorithm~\cite{Cohen03:BitTorrent} has equalized the replication degree of different chunks at the various neighborhoods. From that point in time on, we can factor out chunk availability and estimate the established unchokes based only on the uplink speed of nodes. In Appendix~\ref{appendix:bmatching_completion} we extend this numeric method to capture also the initial flash-crowd phase of a torrent. The resulting model is much slower in terms of execution time and provides rather limited additional fidelity since the flash crowd phase is known to be relatively short compared to the steady-state phase of sufficiently large downloads (the size of a movie or a software package)~\cite{biersackbittorrent,Legout06:LRFenough,Laoutaris2008:bitmax}. For this reason we stick to the original more scalable model.

Notice that although experimentation with real clients would provide higher accuracy in predicting the QoS of individual clients, it wouldn't be able to scale to the number of torrents and clients needed for studying the impact of realistic torrent demographics at the ISP level (aggregate traffic in the order of several Gbps). Our scalable numeric methodology targets exactly that while preserving key BitTorrent properties like leecher unchoking (regular and optimistic) and seeding. We validate the accuracy of our methods against real BitTorrent clients in controlled emulation environments (Appendix~\ref{appendix:validation}) and in the wild with live torrents (Sect.~\ref{sec:prototype}).

\subsection{Modeling Seeders}\label{subsec:modelingseeders}

Let $N(s,T)\subseteq$ be the neighborhood of a seeder node $s$ of torrent $T$. Existing seeders typically split their uplink capacity $U(s)$ among their neighbors following one of two possible policies. In the \emph{Uniform} policy, all neighbors $u\in N(s,T)$ get an equal share $upload(s,u)=U(s)/|V(s,T)|$. In the \emph{Proportional} policy, neighbor $u\in N(s,T)$ gets an allotment in proportion to its speed, \ie, $upload(s,u)=U(s)U(u)|/\sum_{u'\in N(s,T)} U(u')$.

\subsection{Modeling Leechers}\label{subsec:modelingleechers}

Estimating the traffic flow among leechers is more involved due to the unchoke algorithm~\cite{Cohen03:BitTorrent}. This reciprocity based matching algorithm of nodes with similar speeds has many of the elements of a $b$-matching problem~\cite{Cechlarova05:bmatching,gai07:bmatching}. In Appendix~\ref{app:bmatching_basic} we show how to cast the problem of estimating the traffic matrix from a torrent $T$ downloaded by nodes in $V(T)$ as a $b$-matching problem in $V(T)$. We also point to work describing how to get a fast solution (a stable matching $M$) for the $b$-matching. $M$ gives us the pairs of nodes that unchoke each other in steady-state. Using the stable matching $M$ and the uplink
speeds of nodes, we can compute the expected rate at which a
node $v$ uploads to its neighbor $u$:

\[
\small
upload(v,u) = \left \{ \begin{array}{ll}
\frac{U(v)}{k+1}, & \qquad \textrm{if } (v,u)\in M\\
\frac{U(v)}{k+1} \cdot \frac{1}{|N(v,T)|-k}, & \qquad
\textrm{otherwise}
\end{array} \right.
\]

The first case amounts to neighbors $u$ that are allocated one of
$v$'s $k$ regular unchokes in steady-state. The second case
amounts to the remaining neighbors that receive only optimistic
unchokes and thus share the single slot that is allocated
optimistically.\footnote{\footnotesize It might be the case that
in a stable solution node $v$ is matched to less than $k$ others
(e.g., because it is of low preference to its neighbors). In such
cases we add the unallocated unchoke bandwidth to the optimistic
unchoke bandwidth that is evenly spread to choked neighbors.} In
Appendix~\ref{appendix:validation} we have validate the accuracy
of the $b$-matching for estimating unchokes during the two typical
phases of a torrent's lifetime~\cite{Pouwelse05:BitTorrent}. In
Sect.~\ref{sec:understanding} we will use the upload rates from
the $b$-matching described above and the corresponding seeder
bandwidth allocation policies of
Sect.~\ref{subsec:modelingseeders} to compute the amount of
BitTorrent traffic crossing inter-AS links. Before that, however,
we introduce the overlay construction policies that we will study.

%% file: overlayconstruction.tex
\section{Locality-biased Overlays}\label{sec:overlayconstruction}

Up to now our discussion has been based on a very basic locality biasing overlay construction algorithm, Locality, that provides a node $v$ of $A$ participating in $T$ with $\min(W,|V(A,T)|-1)$ local nodes and pads up to $W$ with randomly chosen remote nodes. In this section we want to generalize Locality so that it can capture the operation of existing overlay construction policies like the ones proposed in~\cite{Xie08:P4P,choffnes08:ONO}.

\subsection{A family of locality-biased overlays}

We refer to the resulting extended family of overlay construction algorithms as Locality($\delta,\mu$). Its operation is as follows. It starts with a neighborhood $N(v,T)$ of $\max(W,|V(T)|-1)$ randomly selected neighbors which are then filtered based on speed comparisons against the set of local nodes $V(A,T) \backslash \{v\}$. These comparisons are modulated by the parameters $\delta,\mu$ as follows. Parameter $\mu$ controls the maximum number of allowed remote (inter-AS) neighbors in $N(v,T)$. If the number of remote nodes in $N(v,T)$ is greater than $\mu$ then a remote node $u$ is substituted by a local $w$ that is not already in the neighborhood until the number of remotes reaches $\mu$. If there are no more
local nodes for performing switches then $u$ is taken out of the neighborhood. If the number of remotes in $N(v,T)$ is already below $\mu$, then $u$ is substituted by a not already taken local node $w$ only if $1-\frac{U(w)}{U(u)}<\delta$.

For now we won't concern ourselves with implementation issues as we only use these policies for exploring the transit reduction vs. user QoS tradeoff. We will discuss implementation issues later in Sect.~\ref{sec:prototype}.

\subsection{Some notable members}

In the evaluation presented in Sect.~\ref{sec:understanding} we will consider some members of the Locality($\delta,\mu$) that are of special interest. These include:

\begin{itemize}

 \item $\delta = 0, \mu=\min(W,|V(T)|-1)$: In this case there is no constraint on the number of remote neighbors whereas switches of remote for local nodes occur only if the local ones are faster. We call this end-user QoS preserving policy \emph{Local Only If Faster (LOIF)}.
 \item $\delta = 1, \mu=\min(W,|V(T)|-1)$: Again there is no constraint on the number of remote neighbors but local nodes are preferred independently of their speed comparison to remotes. This is the standard \emph{Locality} we introduced in Sect.~\ref{sec:implicit}.
 \item $\delta = 1, \mu=1$: As before all switches of remotes for locals are performed. Of the remaining remotes only one is retained and the rest are discarded from the neighborhood. We call this policy \emph{Strict}.

\end{itemize}

%% file: understanding.tex
\section{Impact of Locality on ISPs \& Users}\label{sec:understanding}

The bounds presented in Sect.~\ref{sec:implicit} provide a
broad view of the impact of locality on the transit traffic of
hundreds of ISPs. They do not, however, provide any information
regarding the impact of locality on end user download rates. Earlier
work~\cite{Xie08:P4P,choffnes08:ONO} has demonstrated some
``win-win'' cases in which ISPs benefit by reducing their transit
traffic, while at the same time their users get faster download
rates. In general, however, talking mostly to local nodes can harm
a user's download rate by, \eg, depriving it from faster remote
seeders and leechers (the latter can provide optimistic unchokes).
Whether this happens depends on the interplay between demographics
and speed. In this section we employ the traffic matrix
computation methodology of Sect.~\ref{sec:bittorrenttrafficmatrix}
to present detailed case studies of the impact of different
overlay construction mechanisms from
Sect.~\ref{sec:overlayconstruction} on ISPs and their users.
\emph{We are primarily interested in discovering the boundaries of
the win-win region from locality for both ISPs and users as well
as the reasons behind them}.

\subsection{Experimental methodology}

Next we present the common parts of our methodology that appear in
all experiments. Experiment-specific parts appear in the corresponding
sections.

\subsubsection{Input to the experiments}\label{subsec:inputtoexp}

\vspace{2pt}

\noindent \textbf{Demographics:} We used the BitTorrent demand
demographics measurements presented in
Sect.~\ref{sec:demographics}. If not otherwise stated, our default
dataset will be ${\tt mn40K}$.

\vspace{2pt}

\noindent \textbf{Speed distributions:} If not otherwise stated, we
assign to an ISP the median uplink speed of its country~\cite{Stanford:page}. We also use speeds from
iPlane~\cite{iplane_osdi}. One important point is that these
represent last mile bottlenecks. We consider network bottlenecks later in Sect.~\ref{sec:prototype}
using an experimental prototype and live torrents.

\vspace{2pt}

\noindent \textbf{Seeder/leecher ratios:} In dataset {\tt pb600}
we know exactly if a peer is seeder or leacher but in {\tt mn40K}
and {\tt mn3K} we do not have this information. To solve this
problem, we obtained from the correspondent tracker the number of
seeders and leechers for each torrent. Then we made a client in
our dataset a seeder with probability equal to the seeder/leecher
ratio of its torrent. Thus although we don't have the exact
identities of seeders, we do match the real seeder/leecher ratios.
We validated this technique with the dataset {\tt pb600} obtaining
minor variation compared to real seeder distributions.
The raason for this is that the seeder/leecher ratio is fairly stable across ISPs.

%we focused on a limited number oftorrents for which we identified the real seeders using bitfields.Analyzing these torrents we saw that the seeder/leacher ratio isfairly stable across ASes, \ie, most ASes seem to be seeding with equal intensity. This reduces the noise introduced by our seeder allocation method.

\begin{table*}[tb]

\begin{center}
\subtable[\small Transit traffic reduction under {\tt mn40K} and
Ookla speeds.]{
%\centering
%\small
\begin{tabular}{l|l|l|l|}
ISP  & LOIF & Locality & Strict \\  \hline
\bf{US1}  & \bf{32.00\%} & \bf{55.63\%} & \bf{97.47\%} \\
US2 & 28.47\% & 48.40\% &97.25\% \\
US3 & 26.04\% & 41.45\% & 97.02\% \\
\bf{EU1}  & \bf{10.50\%} & \bf{39.12\%} & \bf{96.41\%} \\
EU2 & 11.34\% & 44.89\% &95.95\% \\
EU3 & 16.18\% & 35.57\% & 96.98\% \\
\end{tabular}
%\caption{Transit traffic reduction under {\tt mn40K} and Ookla speeds.}
\label{tab:40k-tt} }\hspace{30pt} \subtable[\small Degradation of median QoS
under {\tt mn40K} and Ookla speeds.]{
%\centering
%\small
\begin{tabular}{l|l|l|l|}
ISP  & LOIF & Locality & Strict \\  \hline
\bf{US1} & \bf{-6.71\%} & \bf{-1.32\%} & \bf{2.88\%} \\
US2 & -5.22\% & -0.83\% & 4.43\% \\
US3 & -5.74\% & -1.27\% & 4.96\% \\
\bf{EU1} & \bf{-1.47\%} & \bf{3.33\%} & \bf{18.59\%} \\
EU2 & -0.55\% & 6.35\% & 11.72\% \\
EU3 & -3.21\% & 2.28\% &14.67\% \\
\end{tabular}
\label{tab:40k-qos} }

\vspace{0.01pt}

\subtable[\small Transit traffic reduction under {\tt pb600} and
Ookla speeds.]{
%\centering
%\small
\begin{tabular}{l|l|l|l|}
ISP  & LOIF & Locality & Strict \\  \hline
\bf{US1} & \bf{34.03\%} & \bf{77.86\%} & \bf{99.10\%} \\
US2 & 30.56\% & 69.20\% & 98.73\% \\
US3 & 37.11\% & 78.70\% & 99.27\% \\
\bf{EU1} & \bf{15.25\%} & \bf{72.80\%} & \bf{99.35\%} \\
EU2 & 21.22\% & 72.26\% &99.18\% \\
EU3 & 26.57\% & 71.92\% & 99.05\% \\
\end{tabular}
\label{tab:top600-tt} }\hspace{30pt} \subtable[\small Transit
Traffic Reduction under {\tt mn40K} and iPlane speeds.]{
%\centering
%\small
\begin{tabular}{l|l|l|l|}
ISP  & LOIF & Locality & Strict \\  \hline
\bf{US1} & \bf{16.14\%} & \bf{52.12\%} & \bf{96.63\%}  \\
US2 & 8.77\%  & 46.73\%  & 95.68\% \\
US3 & 9.18\%  & 39.55\%  & 94.66\% \\
\bf{EU1} & \bf{3.94\%}  & \bf{43.89\%}  & \bf{94.92\%}  \\
EU2 & 5.68\%  & 50.89\%  & 94.69\% \\
EU3 & 12.68\% & 41.63\%  & 95.62\% \\
\end{tabular}
%\end{center}
%\caption{Transit Traffic Reduction under {\tt mn40K} and iPlanespeeds.}
\label{tab:40k-tt-iplane} }

\vspace{-4pt}
\caption{Results for ISPs EU1-EU3, US1-US3, under different
demographic and speed datasets}
\end{center}
\end{table*}
\vspace{-5pt}

\subsubsection{Traffic matrix computation}

In our experiments we are interested in quantifying the effects of
locality biased overlay construction on a ``home'' ISP $A$. We
perform this as follows.

\vspace{3pt}

\noindent (1) Using our measured demand demographics we identify
the set of clients $V(T)$ for each torrent $T\in T(A)$ downloaded
by clients in our home ISP $A$. We construct Random, LOIF,
Locality, and Strict overlay graphs among the nodes in $V(T)$ as
described in Sect.~\ref{sec:overlayconstruction}. We select the
nodes to be seeders as described in Sect.~\ref{subsec:inputtoexp} and assume that they
perform proportional seeding.

\vspace{3pt}

\noindent (2) We feed each overlay graph resulting from the
combination of the demographics of a torrent $T$ and an overlay
construction algorithm, together with an uplink speed distribution
to the BitTorrent traffic matrix computation methodology detailed
in Sect.~\ref{sec:bittorrenttrafficmatrix}. The outcome is a
traffic matrix indicating the transmission rate between any two
nodes $v,u \in V(T)$.

\vspace{3pt}

\noindent (3) We adopt a simplified version of routing according to which all
traffic between clients of our home ISP and an ISP of the same
country goes over unpaid peering links, whereas traffic between
clients of our home ISP and another ISP in a different country goes over
a paid transit link. This simplified routing is actually on the conservative side, since it reduces the amount of traffic going to the transit link and thus also the potential gains from applying locality.

\vspace{3pt}

Repeating steps (1)--(3) for all torrents in $T(A)$ we obtain the
aggregate amount of traffic going to the transit link of $A$ due
to the torrents appearing in our dataset.

\subsubsection{Performance metrics}

We study two performance metrics. The first one, \emph{transit
traffic reduction compared to random} is of interest to the home
ISP. It is defined as follows: ( (aggregate transit under
Random)-(aggregate transit under Locality($\delta,\mu$)) ) /
(aggregate transit under Random). The second one, \emph{user QoS
reduction} is of interest to the clients of the home ISP. It is defined
as follows: ( $q_x$(download rate under Random)-$q_x$(download rate under
Locality($\delta,\mu$)) ) / $q_x$(download rate under Random), where $q_x$ denotes the $x$-percentile of download rate computed over all nodes of home ISP. If not otherwise stated we will use the median ($x=0.5$).

\subsection{Comparing overlays}\label{subsec:transitqos}

In Table~\ref{tab:40k-tt} we present the transit traffic reduction
under various locality policies with respect to Random for the 6
largest ISPs (3 from Europe and 3 from US) across our different datasets using
uplink speeds from~\cite{Stanford:page}. In
Tables~\ref{tab:40k-qos} we present the corresponding impact on
user QoS. We have obtained similar results for several other ISPs.
We will comment mainly based on the ISPs, EU1 and US1, introduced
earlier in Sect.~\ref{subsubsection:casestudy1}.

\subsubsection{Without bounding the number of inter-AS links}

We begin with ``mild'' locality policies that do not enforce
constraints on the number of remote neighbors. The mildest of all,
LOIF, replaces remote with local nodes in the neighborhood only if
the locals are faster. In the case of US1 this yields a transit
traffic reduction of 32\% compared to Random. The corresponding
value for EU1 is 10.5\%. US1 is faster than EU1 and performs more
switches of remotes for locals under LOIF and thus gets a higher
reduction of transit traffic. Looking at Table~\ref{tab:40k-qos}
we see that US1 pays no penalty in terms of QoS reduction for the
end users from LOIF. Actually, the median value gets a slight speedup indicated
by negative values (see Appendix~\ref{app:otherpercentiles} for other percentiles). The situation for EU1 is similar. The preservation of at least the same user QoS is an inherent
characteristic of LOIF which by default leads to a win-win
situation for both ISPs and users. The transit savings of LOIF can
however be small, as in the case of EU1.

We can reduce the transit traffic further by imposing a less
strict switching rule. Locality switches any remote client with a
local one independently of speed. This increases the savings for
US1 to 55.63\% compared to Random whereas the corresponding number for EU1
rises to 39.12\%. This is the highest transit reduction that can
be expected without limiting the number of inter-AS overlay links.
This additional transit traffic reduction does not impose any QoS
penalty on the customers of US1. EU1 customers on the other hand
pay a small reduction of QoS of 3.33\% since they loose
some faster remote neighbors (EU1 is not among the fastest ISPs
according to the country speeds depicted in
Fig.~\ref{fig:cdf_country_speed}). Under Locality win-win is not
guaranteed but rather it depends on speed and demographics. For
US1 Locality is again a clear win-win whereas for EU1 is almost
win-win.

\subsubsection{Unlocalizable torrents}\label{subsubsec:unlocalizable}

In the aforementioned results the transit traffic reduction toped
at around 55\%. This happens because the demographics of both US1
and EU1 include a long tail of torrents with very few local nodes.
These torrents are ``unlocalizable'' in the sense that all overlay
links for them will have to cross the transit link if the
corresponding clients are to be given the standard number of
neighbors according to BitTorrent's bootstrapping process (40-50
depending on version). The unlocalizable torrents put rigid limits
on the transit reduction that can be achieved without enforcing
constraints on the number of allowed inter-AS overlay links.
Interesting, although the unlocalizable torrents create most of the transit traffic, they are
requested by a rather small percentage of the nodes of an ISP. In
US1 90\% of transit traffic under Locality is due to only 10\% of
the nodes. In EU1 90\% of transit traffic is due to 13.44\% of nodes.

\subsubsection{Bounding the number on inter-AS overlay links}

If we want further transit traffic reductions then we need to
control the unlocalizable torrents by enforcing strict constraints
on the number of inter-AS overlay links. In the last column of
Table~\ref{tab:40k-tt} we depict the performance of Strict that
permits up to 1 inter-AS overlay link per torrent for a given
client. Indeed in this case the transit traffic reduction is huge
(around 96\%-97\% for both networks). The median user QoS drops by 18.59\% in EU1. The
situation is much better for US1 where the median speed drops by around 3\%. However, nodes that are downloading unlocalizable torrents pay a heavy penalty of almost 99\%.

\subsection{Comparing ISPs}

Inspecting Table~\ref{tab:40k-tt} we see that American ISPs in
general achieve higher transit traffic reduction than European
ones, across all locality biased overlay construction policies. We
attribute this to the fact that Random performs very poor in those
ISPs since their content is more scattered around the world
(they have smaller Inherent Localizability,
Sect.~\ref{subsubsection:casestudy1}). When comparing among
American or among European ISPs, the observed differences
correlate mostly with the size of the ISP. The reason is that in
ISPs with approximately the same Inherent Localizability (\eg, the 3 American ISPs), Random
performs approximately the same, and thus any difference in
transit reduction comes from the performance of Locality or LOIF. The latter depend on the absolute
size of the ISP since a larger ISP can gather more easily enough
local peers to reach the minimum number required by the
bootstrapping process.

%% file: prototype.tex
\vspace{10pt}

\section{Validation on live torrents}\label{sec:prototype}

In our study up to now, we have only considered last mile
bottlenecks imposed by access link speeds but no network
bottlenecks due to congestion or ISP traffic engineering,
including throttling~\cite{Marcel08:bittorrentblocking}. In
addition, although we have evaluated the accuracy of $b$-matching
in a controlled emulation environment
(Appendix~\ref{appendix:validation}), we can obtain further
insights by testing our results in the wild where we can observe
additional effects from delays and losses that are lacking from an
emulation environment that captures only bandwidths. To address
such issues we integrated LOIF, Locality, and Strict($\mu$) into the mainline Bittorrent client
(version 5.2.2). Next we
describe briefly some implementation issues and then move to
present the results of connecting to live
torrents with our modified client.

\subsection{Prototype implementation}

Integrating these policies into the existing BitTorrent clients requires addressing some new requirements.
First, we need to know for every Bittorrent client its ISP and country. For this, we use the MaxMind geolocation database~\cite{maxmind:page}. Next, we need to discover the list of local
clients in order to be able to substitute remote ones with local ones. For this, we use the PEX messages
to discover all the participants of the swarm, and then use the MaxMind database to classify them.
Finally, the last requirement which is specific to LOIF, is to estimate the speed of the local
and remote clients. For this, we monitor the rate at which the clients send us HAVE messages, which indicates
how fast they download. Finally, notice that the above method works only for inferring the speed of
leechers. For seeders, we can only infer the speed of those seeders that unchoke us. Thus in LOIF we
do not switch neighbors for which we do not have speed information.

Next, we briefly describe our implementation of LOIF. Many
implementation decisions are influenced by how the python mainline
Bittorrent client is designed and implemented. Every 40 seconds we
perform the following steps:

\begin{itemize}
\item  We classify the active neighbor peers in 3 lists: $L_{ISP}$ which contains all peers that belong to the same ISP as the client, $L_{Peering}$ which contains all peers that are in ISPs with peering relationships and $L_{Remote}$ which contains other peers.

\item For every peer $R_i \in L_{Remote}$, we close the connection if there exists a peer $L_j \in L_{ISP}$ with higher estimated download speed. If such a peer does not exist then we check if there exists a peer $C_j \in L_{Peering}$ with higher estimated download speed and in this case again we close the connection.\footnote{The assumption is that nodes in the same country communicate over peering links. In our implementation we do not infer ISP relationships but we can do so with iPlane Nano~\cite{Madhyastha09:iplanenano}.}

\item For each connection that was closed in the last step, the algorithm opens a new one, giving
preference, first to those IPs that belong to the same ISP, then to those IPs belonging to peering ISPs and, finally, to those IPs belonging to other ISPs.

\end{itemize}

An important detail in our implementation is to always have a minimum number of neighbors (at least 40).
This holds for LOIF and Locality, but not for Strict.
% of at  stable  number of number With LOIF and Locality we always tried to provide the standard number of neighbors
 For Strict($\mu$), we close connections and don't open new ones, if we have more than $\mu$ remote nodes.

\begin{figure}[tb]
\centering
%\vspace{1.5cm}
\includegraphics[width=1.6in, angle=-90]{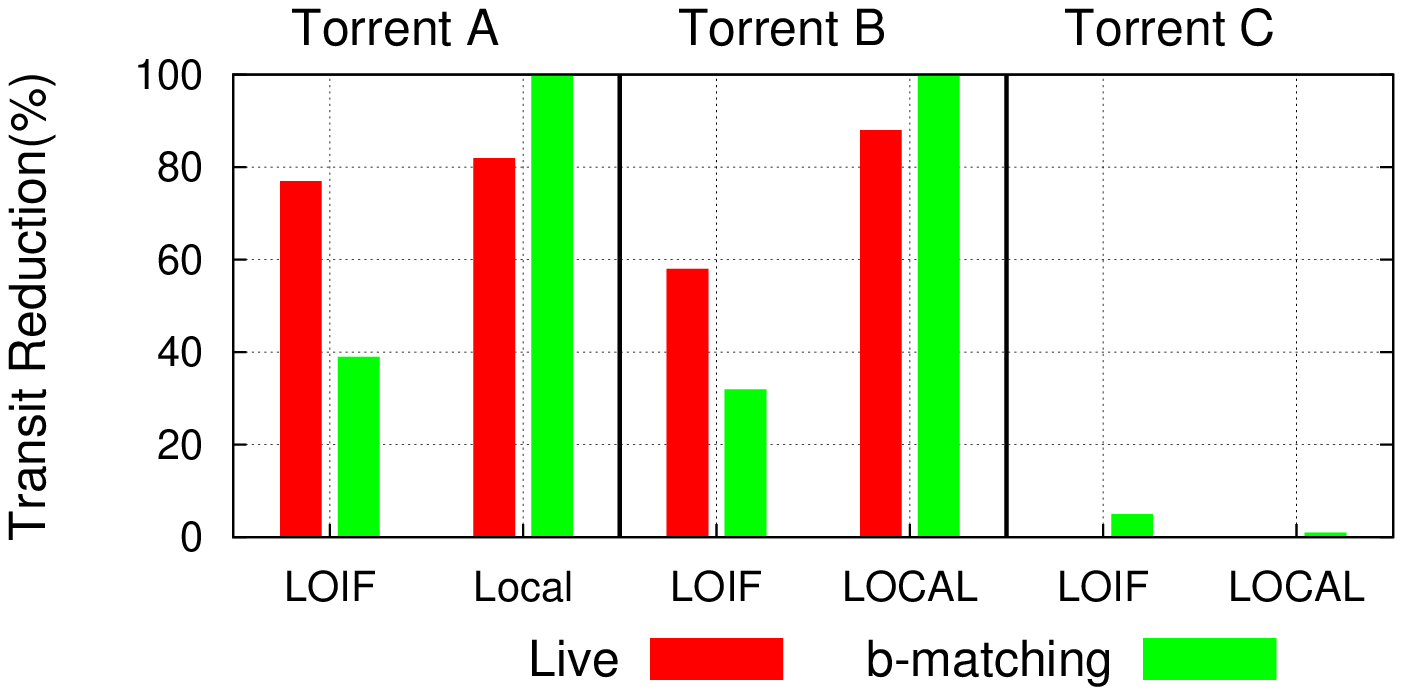}
\caption{Comparision between LOIF and Local}
\label{fig:slow_isp_medium_content}
\vspace{-5pt}
\end{figure}

\begin{table}[tb]
\begin{center}
\begin{tabular}{cccc}
\hline
 & Local & Remote & Percentage of Seed \\  \hline
Torrent A & 521 & 46 & 63.2\% \\ %\hline
Torrent B & 351 & 211 & 12.5\% \\ %\hline
Torrent C & 3 & 666 & 66.4\% \\ \hline
\end{tabular}
\end{center}
\vspace{-5pt}
\caption{Live torrent characteristics}
\label{tab:torrent_characteristics}
\end{table}
\vspace{-7pt}

\subsection{Experimental methodology}

We ran our modified Bittorrent client from an ADSL connection of
ISP EU1. In all the torrents we first warmed up by downloading
30MB to avoid BitTorrent's startup phase. In each run, we
re-initialize back to the same 30MB. Next, we download 50MB with
each of the following policies: Random, LOIF, Locality, and
Strict. We repeated each test 5 times, and reported averages over
all runs. During each experiment we logged the list of IPs and the
number of neighbors and used them later as input to our traffic
matrix estimation technique of
Sect.~\ref{sec:bittorrenttrafficmatrix}. This way, we can compare
the estimated transit savings with the real one on live torrents.

\subsection{High, medium, and low localizability torrents}

We used our prototype to validate some of our previous results. Although we cannot scale to the number of torrents discussed in Sect.~\ref{sec:understanding}, we tested torrents at characteristic points of the demographic spectrum. In particular, we tested a very popular torrent inside EU1 (Torrent A), an intermediately popular one (Torrent B), and an unpopular one (Torrent C). In Table~\ref{tab:torrent_characteristics} we summarize the characteristics of the 3 torrents. In Fig.~\ref{fig:slow_isp_medium_content} we present the transit traffic savings as predicted by our traffic matrix estimation method and as measured on the live torrent under LOIF and Locality. We do not present results under Strict as they were always in perfect agreement.

Overall we see that the results under Locality are pretty consistent -- estimation and measurement are within 10-20\% of each other. In terms of absolute values things are as expected: in cases A and B there are enough local nodes to eliminate almost all transit traffic whereas in C there is 0 saving as there do not exist any local nodes to switch to. The difference between the 100\% savings predicted by $b$-matching in A and B and the ones measured in practice has to do with implementation restrictions. As mentioned earlier, we update the overlay every 40~$sec$ (which is equal to 4 unchoke intervals). During that time new incoming remote connections are accepted and can lead to unchokes that create transit traffic and thus eat away from the 100\% saving expected upon overlay update instants when all remote connections are switched with local ones.

Under LOIF, the deviation between estimation and measurement is substantial: the measured transit saving is twice as big as the estimated one. To interpret this, we looked at the number of switches of remote nodes for local ones that LOIF performed in practice and realized that they were much more than we would predict. This in effect means that the real LOIF found the remote nodes to be slower than what expected from our speed dataset from Ookla~\cite{Stanford:page}. We attribute this to network bottlenecks or throttling at the inter-AS links of EU1 or the ISPs that host the remote nodes. Although certainly interesting, identifying exactly why the remote nodes appear slower than expected is beyond the scope of the current work. See~\cite{Marcel08:bittorrentblocking} for more.

%% file: related.tex
\section{Related Work}\label{sec:related}

\subsection{Early work on locality-biasing}

One of the early works on locality-biased overlay construction was Karagiannis et al.~\cite{Karagiannis05:P2PIMC}. Using
traces from a campus network as well as a six-month-long logfile
from a popular torrent, they showed that there is substantial overlap in the torrents downloaded by
co-located clients. Another early work from Bindal et
al.~\cite{Bindal06:BitTorrent}, studied the effect of limiting the
number of inter-AS connections using simulations with synthetic
demand. Aggarwal et al.~\cite{Aggarwal07:p2poracle} studied the
effects of locality biasing on the Gnutella overlay. Apart from
studying a different P2P system, they differ from our work by
focusing on the overlay graph theoretic properties whereas we care
about the traffic matrix.

\subsection{Recent systems for locality-biasing}

Following up on positive results on the potential of
locality-biasing, a number of actual systems like
P4P~\cite{Xie08:P4P} and ONO~\cite{choffnes08:ONO} have appeared
recently for the BitTorrent P2P protocol. The previous works focus on
architectural and systems questions regarding ``how'' to implement
locality-biasing, and in particular whether the goal can be
achieved through unilateral client-only solutions, or bilateral
cooperation is essential for making locality work for both ISPs
and users. In terms of reported results, \cite{Xie08:P4P} presents
a variety of use cases for P4P over different networks and P2P
applications like Pando and Liveswarms. In all cases however,
results are based on one or a few swarms and thus do not capture
the aggregate effects created by tens of thousands of concurrent
swarms with radically different demographics. The results reported
in~\cite{choffnes08:ONO} on the other hand, are indeed from
multiple torrents and networks, but they only report on the final
outcome from using the ONO system without explaining how the
demographics of the torrents and the speeds of the ISPs affect
these outcomes. The main driving force behind our work is to
explain ``when'' locality works and ``why'' when it does so and
thus help in interpreting the results from systems like P4P and
ONO or others to come in the future. Locality biasing has
also been applied to P2P streaming
systems~\cite{Picconi09:locality}.

\subsection{BitTorrent measurements}

A substantial amount of work has gone into BitTorrent
measurements~\cite{biersackbittorrent,Guo05:BitTorrentMeasurement,Pouwelse05:BitTorrent,Menasche09:bundling}.
These works go beyond locality to characterize things like the
arrival pattern of new nodes, the seeding duration, the
seeder/leecher ratios, etc. Our work apart from performing large
scale measurements develops scalable methodologies that permit
distilling non-trivial conclusions regarding the interplay of
demographics, speed, and overlay construction. Relevant to our work is the recent work of Piatek et
al.~\cite{Piatek09:locality}. It discusses the potential for
win-win outcomes for ISPs and users but puts most of
its emphasis on implementation issues and the
consequences of strategically behaving ISPs. Our work, on the
other hand, is of performance evaluation nature and aims at pushing
the envelope in terms of both the scalability and the fidelity of our evaluation methodology. Our dataset is large; we compute transit reduction from our entire 40K of torrents whereas they use only 1000 torrents out of their 20K dataset. In terms of methodology, we capture the effect of stratification from choke/unchoke
whereas~\cite{Piatek09:locality} assumes cooperative clients and does not model the effect of speed.

%% file: conclusions.tex
\section{Conclusions}\label{sec:conclusions}

In this paper we collected extensive measurements of real BitTorrent demand demographics and developed scalable methodologies for computing their resulting traffic matrix. Based on this we quantified the impacts of different locality-biasing overlay construction algorithms on ISPs and end-users. By studying several real ISPs, we have shown that locality yields win-win situations in most cases. The win-win profile is bounded by ``unlocalizable'' torrents that have few local neighbors. Handling the unlocalizable torrents requires limiting the number of allowed inter-AS overlay connections. This has a small impact on the average user but a dire one on the users of unlocalizable torrents.

%% file: bmatching_basic.tex
\section{Modeling regular unchokes with a $b$-matching}\label{app:bmatching_basic}

%See~\cite{Cuevas2009:LocalityTech}.

The input to a $b$-matching problem consists of a set of nodes $V$,
and functions $n: V\rightarrow 2^V$, $b: V\rightarrow
\mathbb{Z}^+$, and $p: V^2\rightarrow \mathbb{R}^+$ defined as
follows: $n(v)$ defines the set of nodes to which $v$ can be
matched with (matching is symmetric here, and thus $u\in n(v)$ iff
$v\in n(u)$); $b(v)$ defines the number of parallel matchings that
$v$ is allowed to establish; $p(v,u)$ is a measure of the
preference that $v$ has for becoming stably matched to $u$. A
solution to a $b$-matching is a set $M$ of matchings (edges)
between pairs of nodes in $V$, such that for each matched pair
$(v,u)\in M$, the matching and capacity constraints $n,b$ are
satisfied and further, there exists no ``blocking pair''
$(v',u')\in M$, i.e., no pair that satisfies: $p(v,v')>p(v,u)$
and $p(v',v)>p(v',u')$.

It is easy to see that there exists a direct mapping from BitTorrent to
$b$-matching~\cite{gai07:bmatching}. Looking at a particular node
$v$ and torrent $T$: the neighborhood $N(v,T)$ can be mapped to
the allowed matchings $n(v)$; the number of parallel unchokes $k$
(default value for $k$ being 4) at each 10~$sec$ interval
corresponds to $b(v)$, the number of matchings allowed for $v$; the
uplink capacity $U(v)$ of a BitTorrent client $v$ can be used as a
measure of the preference $p(u,v)$ that each node $u\neq v$ would
have for being matched with $v$ in the context of a $b$-matching.
$b$-matchings in which the preference for a node is the same
independently of who is considering, i.e., for given $u$,
$p(v,u)=p(v',u)$, $\forall v,v'$, are said to have a global
preference function. Tan~\cite{Tan91:bmatching} has shown that the
existence of a stable solution for the $b$-matching problem
relates to the non-existence of circles in the preference function
$p$, which is a condition that is certainly satisfied under a
global preference function like $U(v)$. Therefore, for the
aforementioned mapping from BitTorrent to $b$-matching, one can
use a simple $O(|V(T)|\cdot k)$ greedy algorithm to find the
unique stable matching that exists in this
case~\cite{gai07:bmatching}.\footnote{Uniqueness is guaranteed under the assumption that there are not ties in speeds. We made sure that his is the case by adding to our speed datasets a very small random noise.}

%% file: validation.tex
\section{Validation of modeling}\label{appendix:validation}

In this section we validate the accurately of modeling the unchoke algorithm using a $b$-matching. We look at the two typical phases of a torrent's lifetime~\cite{Pouwelse05:BitTorrent}.

\subsection{Startup phase}

\begin{figure}[tb]
\centering
\includegraphics[width=3in]{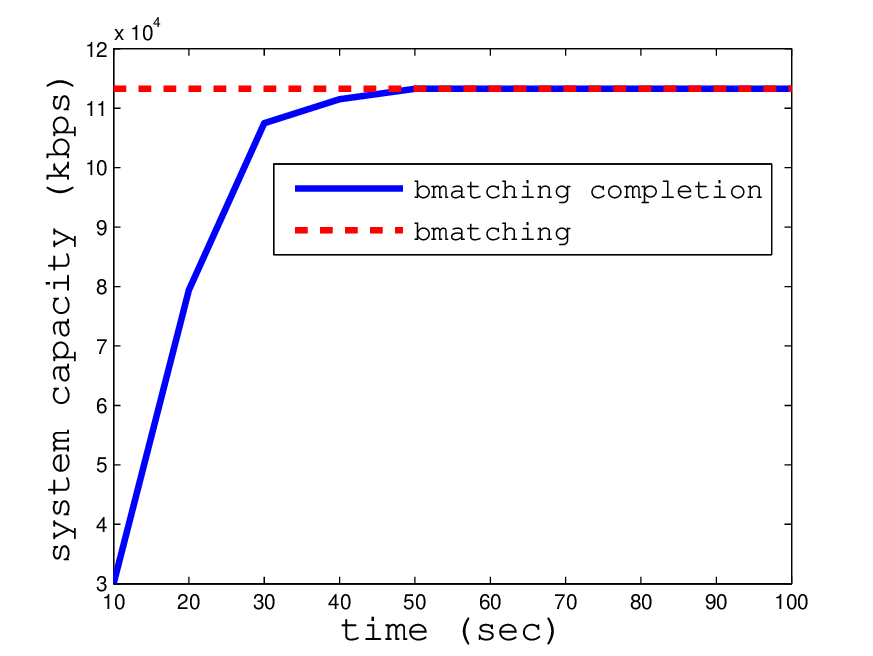}
\caption{\footnotesize Aggregate system capacity from baseline $b$-matching and $b$-matching with completion levels. Parameters: $|V|=40$, uplink rates randomly distributed with mean $2$~Mbps, $C=10000$, $c_0(v)<1$\%, $\forall v$, unchoke duration=10~$sec$, chunk size=32~kBytes. The more complex version converges within a minute to the steady-state value predicted by the baseline $b$-matching. Download completion requires around 30~$mins$.}
\label{fig:test_completion}
\end{figure}

During the initial phase of a new torrent leechers hold few chunks and thus whether two nodes unchoke each other depends, beyond their speeds, on the set of chunks they hold. The $b$-matching modeling of unchoke described in Sect.~\ref{subsec:modelingleechers} assumes that steady-state has been reached and thus chunk (in)availability does not affect the resulting matchings. In Appendix~\ref{appendix:bmatching_completion} we extend this basic matching algorithm to allow it to also capture the \emph{completion level} $c(v)$ of a node, \ie, the percentage of a file of total size $C$ that it holds. We have used this completion-level aware $b$-matching in conjunction with small initial completion levels $c(v)<1\%$ for all leechers to estimate the effect chunk availability on the aggregate capacity of a torrent. BitTorrent's LRF chunk selection strategy is used for expediting the convergence to steady state. We verified this by comparing our two implementations of $b$-matching. In Fig.~\ref{fig:test_completion} we give an indicative example to show that for everything but very small files, the startup phase is much shorter than steady state. For this reason we can ignore it at small cost in terms of accuracy and focus on the baseline $b$-matching that is more scalable to large datasets than the more complicated one involving completion levels.

\begin{figure}[tb]
\centering
\subfigure[$b$-matching Unchoking]{
\hspace{-0.5cm}
\includegraphics[width=1.3in,angle=-90]{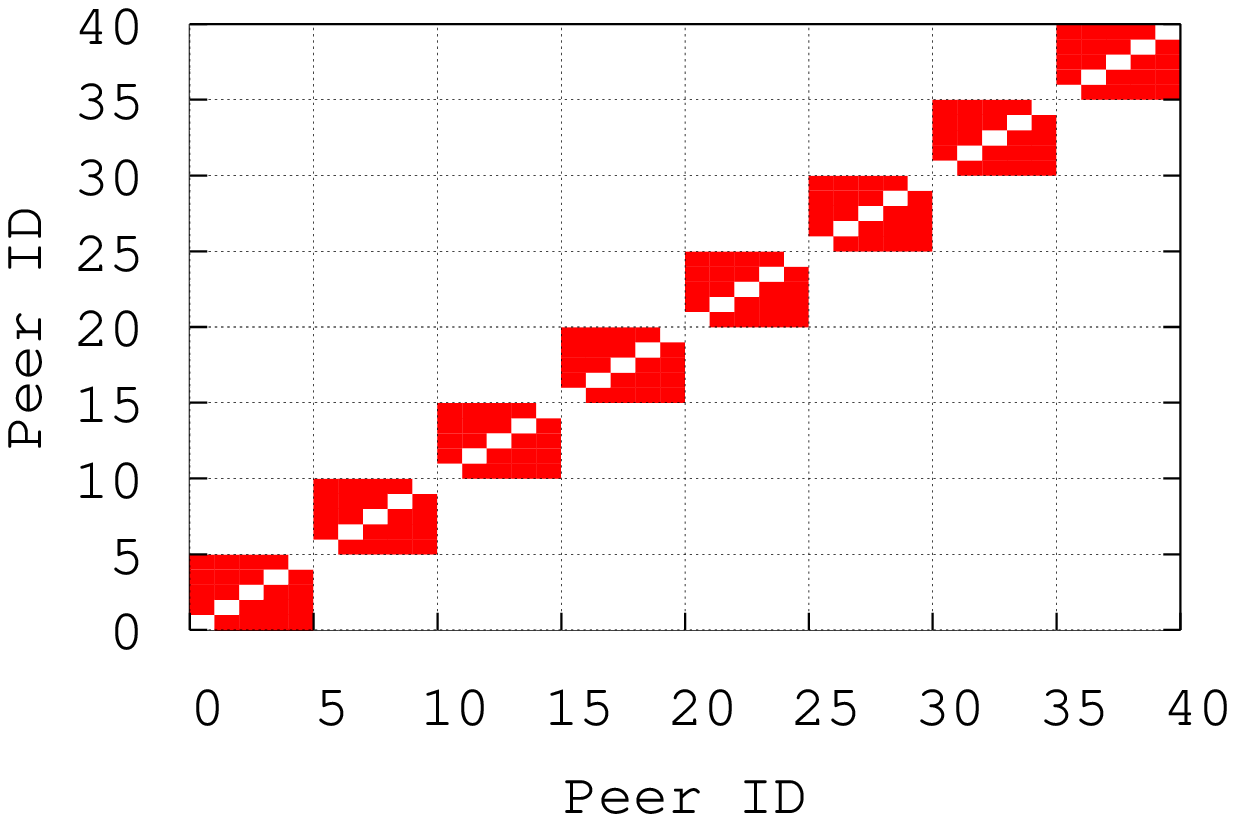}}
\hspace{-1.1cm}
\subfigure[Emulation Unchoking]{
\includegraphics[width=1.3in,,angle=-90]{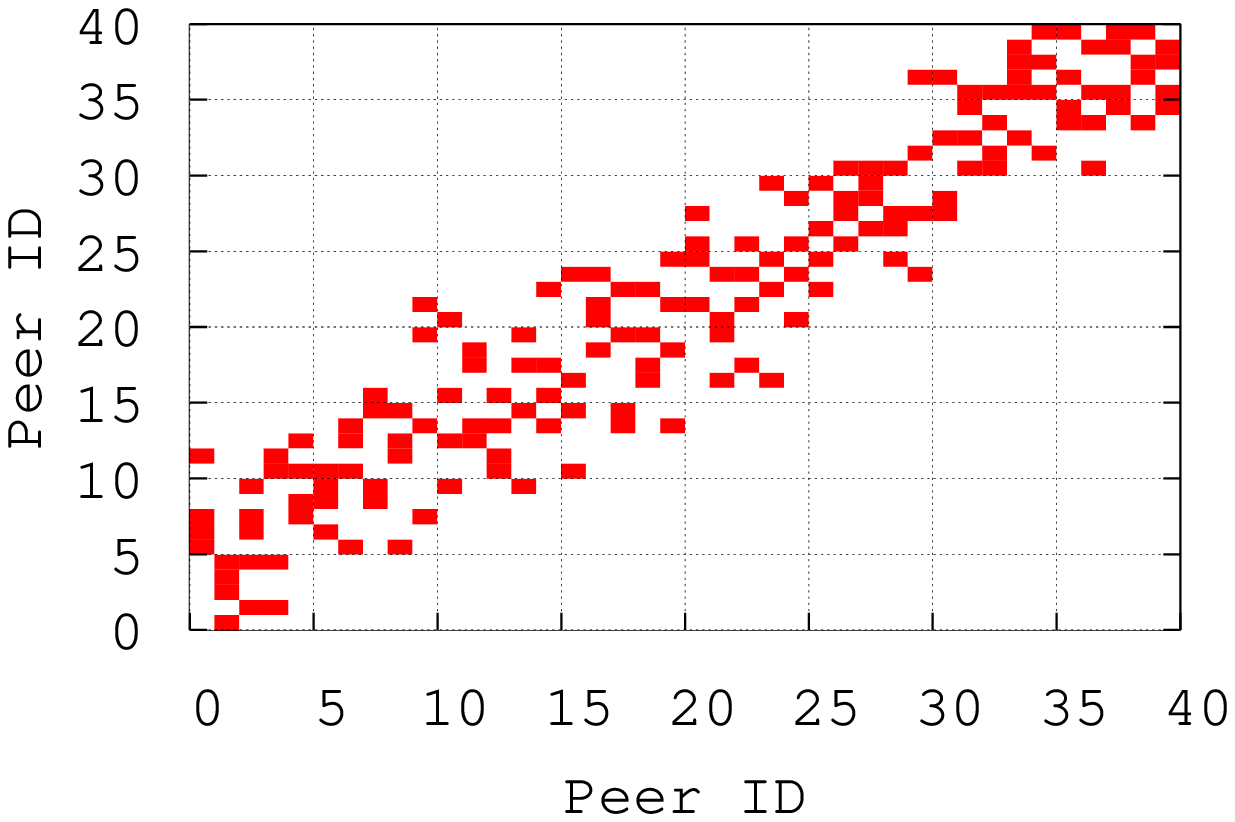}}
\caption{Unchoking Patterns}
\label{fig:bmatching_unc}
\end{figure}

\subsection{Steady state}\label{appendix:modelnet}

Next we validate
the extent at which the steady state matchings predicted by $b$-matching resemble the unchoking behavior of an actual mainline client of BT (v.3.4.2) running in a controlled emulation environment. More
specifically, we set-up a dummy torrent with 40 clients, the
slowest of which, was given an uplink capacity of 80~Kbps, whereas
successive ones were made increasingly faster using a step of
24~Kbps. We chose such a small increment to recreate a rather
difficult environment for stratification~\cite{Legout07:Clustering} to arise. For each
client, we measured the frequency at which the client unchokes
each other client, and then we put a mark on
Fig.~\ref{fig:bmatching_unc}(b) for the $k$ clients that it
unchokes more frequently (client ids on the figure are assigned
according to upload bandwidth in increasing order). Comparing with
Fig.~\ref{fig:bmatching_unc}(a) that depicts the same information
from the execution of $b$-matching under the same uplink
capacities, it is easy to see that $b$-matching provides a
reasonable prediction of the real unchokes that take place and
therefore can be used as a scalable tool for processing huge numbers of small and large torrents that would otherwise be impossible to simulate concurrently. We got similar accuracy using many other torrent sizes
and uplink distributions, including empirical ones from
measurement.

%% file: completion.tex
\section{Completion level aware $b$-matching}\label{appendix:bmatching_completion}

\subsection{Edge filtering}\label{subsec:filtering}

Let $c(v)$ denote the number of chunks already downloaded by node $v$ out of the total $C$ chunks that make up a complete file. For a pair of neighbors $(v,u)$ with $c(v)\geq c(u)$ let $I(v\rightarrow u)$, $c(v)-c(u)\leq I(v\rightarrow u) \leq c(v)$ denote the number of chunks of $v$ that ``are of interest'' to $u$, \ie, chunks that $v$ has downloaded but $u$ hasn't. It is easy to see that $I(u\rightarrow v)=c(u)-c(v)+I(v\rightarrow u)$, $0\leq I(u\rightarrow v) \leq c(u)$. If we assume that the chunks held at some point in time by a node are a random subset of the entire set of chunks, which is reasonable granted LRF~\cite{Cohen03:BitTorrent}, then it follows that:

%{ \small
\begin{equation}
\begin{array}{ll}
p_{vu}(x) & = P\{ I(v\rightarrow u) = x,I(u\rightarrow v) = c(u)-c(v)+x \} \vspace{3pt} \\
& = HyperGeo(c(u)-x,c(v),C,c(u))
\end{array}
\end{equation}
%}

where $HyperGeo(d,p,s,ss)$ denotes a hyper geometric pmf~\cite{Feller1968:probability} giving the probability of drawing $d$ ``successes'' with a sample of size $ss$ from a pool of $p$ items, of which $s$ are ``successes''. Then the expected amount of interest in the two directions is:

\begin{equation}
\begin{array}{ll}
\displaystyle E\{I(v\rightarrow u) \} = \sum_{x=c(v)-c(u)}^{c(v)} x \cdot p_{vu}(x) \vspace{3pt} \\ \displaystyle E\{I(u\rightarrow v) \} = \sum_{x=c(v)-c(u)}^{c(v)} (c(u)-c(v)+x) \cdot p_{vu}(x)
\end{array}
\end{equation}

For pair $(v,u)$ we define its \emph{filtering probability} to be:

\begin{equation}
\phi(v,u)=\min \left( \frac{E\{I(v\rightarrow u)\}}{T\cdot U(v) \cdot (\sigma \cdot k)^{-1}},\frac{E\{I(u\rightarrow v)\}}{T\cdot U(u) \cdot (\sigma \cdot k)^{-1}},1 \right)
\end{equation}

where $\sigma$ is the size of a chunk and $T$ is the duration of an unchoke interval. Given an instance of a $b$-matching problem $\langle V,n,b,p \rangle$ we filter it to obtain a new one $\langle V,n',b,p \rangle$ in which we keep an edge $(v,u)$, meaning that $v\in n(u)$, $u\in n(v)$ and $v\in n'(u)$, $u\in n'(v)$, with probability $\phi(v,u)$, whereas we drop it with probability $1-\phi(v,u)$.

\subsection{Time-evolving completion ratios}

Let $c_t(v)$ be the completion ratio of node $v$ at time $t$ and let $M_t$ be the stable matching obtained from solving the $b$-matching $\langle V,n',b,p \rangle$ in which $n'$ is obtained from $n$ after applying the filtering procedure of Sect.~\ref{subsec:filtering} with completion ratios $\{c_t(v): v\in V\}$. Then the completion ratios of nodes can be updated at the end of the unchoke interval as follows:

\begin{equation}
c_{t+T}(v)=c_t(v)+ \sum_{u: (v,u)\in M_t} \min \left( E\{I(u\rightarrow v)\}, \frac{T\cdot U(u)}{\sigma \cdot k} \right)
\end{equation}

Thus with the above we have a method for mapping the effects of completion levels on the state of a torrent and consequently on the resulting matchings.